\documentclass[10pt, aps, prl, twocolumn,
notitlepage, nofootinbib, longbibliography, showpacs, letter,
superscriptaddress,floatfix
]{revtex4-1}

\usepackage[T1]{fontenc}
\usepackage[latin9]{inputenc}
\usepackage{pstricks,pst-all}
\usepackage{amsmath,bm}
\usepackage{graphicx}
\usepackage{verbatim}
\usepackage[colorlinks=true,urlcolor=blue,linkcolor=red,citecolor=red,breaklinks=true]{hyperref}

\usepackage[normalem]{ulem} 

\begin{document}

\title{Distance-dependent sign-reversal in the Casimir-Lifshitz torque}

\author{Priyadarshini Thiyam}
\email{thiyam@kth.se}
\affiliation{Department of Materials Science and Engineering, Royal Institute of Technology, SE-100 44 Stockholm, Sweden}
\affiliation{Department of Energy and Process Engineering, Norwegian University of Science 
and Technology, NO-7491 Trondheim, Norway}

\author{Prachi Parashar}
\email{prachi.parashar@ntnu.no}
\affiliation{Department of Energy and Process Engineering, Norwegian University of Science 
and Technology, NO-7491 Trondheim, Norway}

\author{K. V. Shajesh}
\email{kvshajesh@gmail.com}
\affiliation{Department of Physics, Southern Illinois University--Carbondale, Carbondale, Illinois 62901, USA}
\affiliation{Department of Energy and Process Engineering, Norwegian University of Science 
and Technology, NO-7491 Trondheim, Norway}

\author{Oleksandr I. Malyi}
\email{oleksandrmalyi@gmail.com}
\affiliation{Centre for Materials Science and Nanotechnology, Department of Physics, University of Oslo, P. O. Box 1048 Blindern, NO-0316 Oslo, Norway}

\author{Mathias Bostr{\"o}m}
\email{Mathias.A.Bostrom@ntnu.no}
\affiliation{Department of Energy and Process Engineering, Norwegian University of Science and Technology, NO-7491 Trondheim, Norway}
\affiliation{Centre for Materials Science and Nanotechnology, Department of Physics, University of Oslo, P. O. Box 1048 Blindern, NO-0316 Oslo, Norway}

\author{Kimball A. Milton}
\email{kmilton@ou.edu}
\affiliation{Homer L. Dodge Department of Physics and Astronomy, University of Oklahoma, Norman, Oklahoma 73019, USA}

\author{Iver Brevik}
\email{iver.h.brevik@ntnu.no}
\affiliation{Department of Energy and Process Engineering, Norwegian University of Science and Technology, NO-7491 Trondheim, Norway}

\author{Clas Persson}
\email{clas.persson@fys.uio.no}
\affiliation{Department of Materials Science and Engineering, Royal Institute of Technology, SE-100 44 Stockholm, Sweden}
\affiliation{Centre for Materials Science and Nanotechnology, Department of Physics, University of Oslo, P. O. Box 1048 Blindern, NO-0316 Oslo, Norway}

\begin{abstract}
The Casimir-Lifshitz torque between two biaxially polarizable anisotropic planar 
slabs is shown to exhibit a non-trivial sign-reversal in its rotational sense. The 
critical distance $a_c$ between the slabs that marks this reversal is characterized 
by the frequency $\omega_c\!\sim \!c/2a_c$ at which the in-planar polarizabilities 
along the two principal axes are equal. The two materials seek to align their principal 
axes of polarizabilities in one direction below $a_c$, while above $a_c$ their axes 
try to align rotated perpendicular relative to their previous minimum energy 
orientation. The sign-reversal disappears in the nonretarded limit. Our perturbative 
result, derived for the case when the differences in the relative polarizabilities are 
small, matches excellently with the exact theory for uniaxial materials. We illustrate our results for black phosphorus and phosphorene. 
\end{abstract}
\maketitle
%
The Casimir-Lifshitz force\,\cite{Casimir:1948dh,*Lifshitz:1956sb,*Dzyaloshinskii:1961fw} between neutral objects in the 
mesoscopic scales has been well established by the modern precision experiments\,\cite{Lamoreaux:1997cfe,*Lamoreaux:1998err,*Mohideen:1998pmc,*Chan:2001mco,*Bressi:2002pcm,*Mohideen:2002lcf,*Decca:2003cfm,*Munday:2009fl,*Chan:2016nmcf}. This force
is a manifestation of the quantum fluctuations in the electromagnetic fields that are 
confined by the boundaries, and is a retarded long-wavelength analogue of the van der Waals force when the finite speed of light $c$ is taken into consideration. A conceptually related 
but a significantly challenging problem is that of the Casimir-Lifshitz torque, which 
arises when the rotational symmetry of the system is disrupted. The 
arduousness of calculating the Casimir-Lifshitz torque, including retardation effects, is demonstrated by 
the fact that the exact analytic evaluation of the torque between two uniaxially anisotropic  semi-infinite half slabs by Barash\,\cite{Barash:1977ba} has never been reproduced by independent methods. A similar evaluation of the torque between two biaxially anisotropic materials is still lacking. 
Barash's calculation is the theoretical basis for the experimentally motivated 
papers in Refs.\,\cite{Munday:2005tbp,*Munday:20085errata,*Romanowsky:2008hac,*Somers:2017rcf,Somers:2017prl}.
The evaluation of the torque does become tractable
in the nonretarded limit\,\cite{Parsegian:1971anr,*Lu:2016ml}; however, the Casimir-Lifshitz torque obtained in this way underestimates the magnitude even at 1\,nm and fails to capture the non-trivial effects 
originating from retardation.

\begin{figure}[t]
\includegraphics[width=7.9cm]{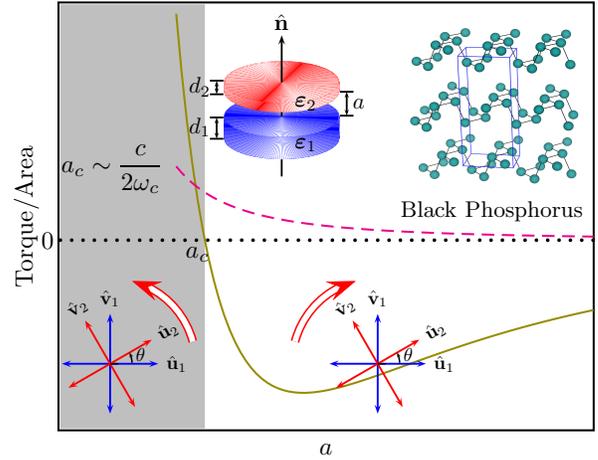}
\caption{
Preview: The torque between two biaxially polarizable planar slabs as a function of the separation distance $a$. Our perturbation theory predicts that when the in-planar polarizabilities along the two principal axes of one of the materials are equal at a characteristic frequency 
$\omega_c$, the torque may change sign at a critical distance $a_c$. Black phosphorus, whose 
crystalline structure is shown in the inset, and its monolayer phosphorene are such materials used in 
our analysis. The solid curve is a preview of our central result, shown here for the torque between two semi-infinite slabs of black phosphorus when the relative angle $\theta=\pi/4$.
The dashed curve is the corresponding nonretarded limit.
}%
\label{tor-main}%
\end{figure}%

In this article, we evaluate the torque between two parallel biaxially anisotropically polarizable slabs of finite thicknesses $d_i$, for $i=1,2$, separated by a distance $a$, as described in the inset 
of Fig.~\ref{tor-main}. We consider the case when the differences in the two in-planar polarizabilities, which are along the two in-planar principal axes, are small. We use a 
perturbative expansion in the parameter 
\begin{equation}
\beta_i(\omega) = \frac{\varepsilon^{u}_i(\omega) -\varepsilon^{v}_i(\omega) }
{\varepsilon^{u}_i(\omega) +\varepsilon^{v}_i(\omega) },
\label{beta}
\end{equation}
that defines the degree of anisotropy in the polarizabilities of two media. Here, $\omega$ 
is the frequency associated with the fluctuations of the fields. 
The leading order perturbative expression for the torque shows a simple linear dependence 
on the product $\beta_1\beta_2$, providing notable qualitative insight. In particular, 
it prompts us to predict a reversal 
in the direction of the torque as a function of the separation distance $a$.  
This change in the
direction of the torque is marked by a critical  distance $a_c\!\sim \!c/2\omega_c$,
where $\omega_c$ is the corresponding characteristic frequency at which the perturbative parameter $\beta_i=0$, i.e., the two in-planar polarizabilities of one media are equal. 
This prediction is motivated by the results in Ref.\,\cite{Elbaum:1991iwr},  
which discusses similar effects in the context of Lifshitz pressure. 

\begin{figure}[t]
\includegraphics[width=8.8 cm]{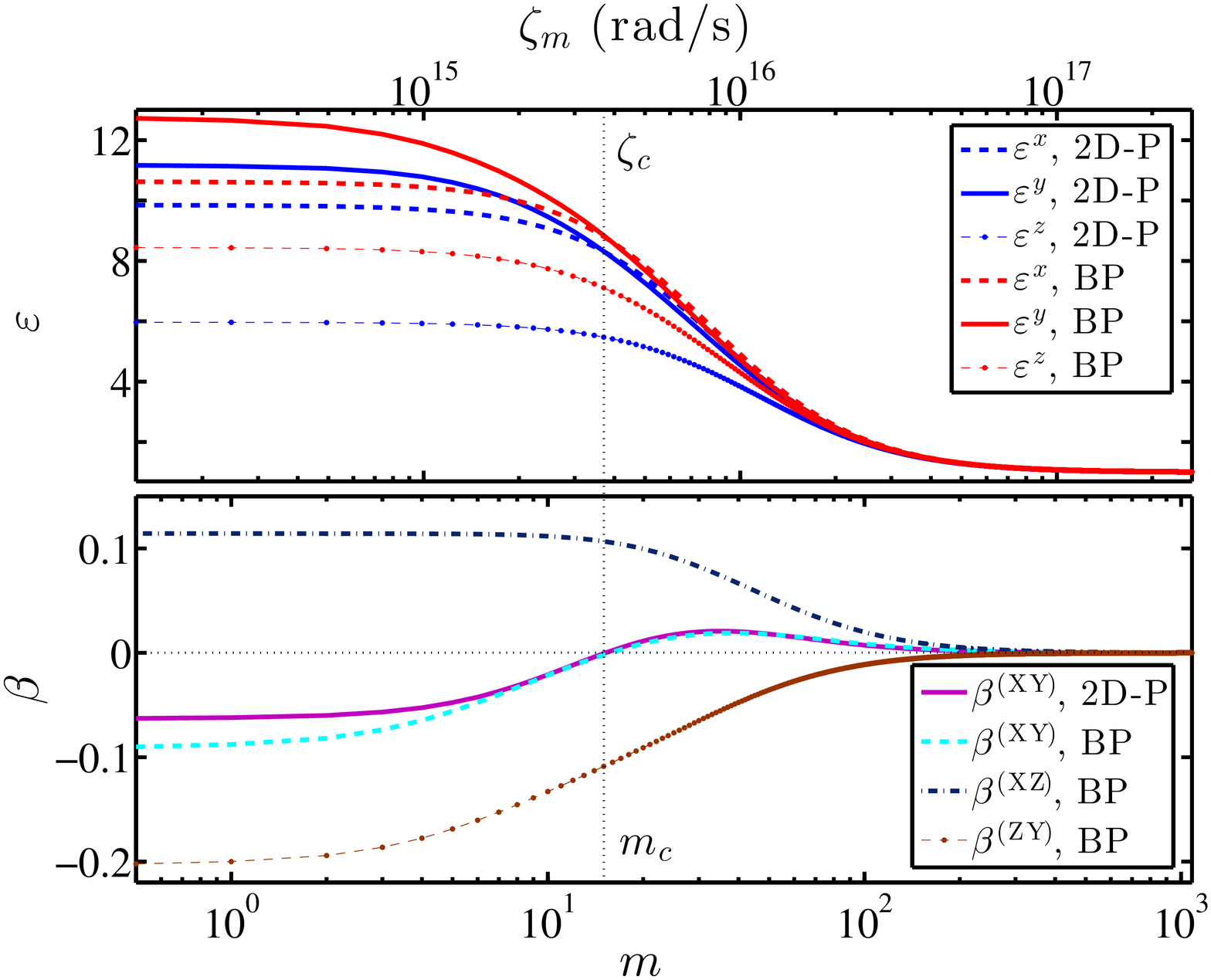}
\caption{Upper panel: Principal components of the dielectric tensors 
${\bm\varepsilon}_i=\text{diag}(\varepsilon^x_i,\varepsilon^y_i,\varepsilon^z_i)$ of $\textrm{BP}$ and $\textrm{2D-P}$, calculated using density functional theory (described in page 3). The imaginary Matsubara frequency $\zeta_m$ is obtained after making the Euclidean rotation;  $\omega_m=i\zeta_m=i 2\pi m k_BT/\hbar$. Components $\varepsilon^{x}_i=\varepsilon^{y}_i$ for both materials close to $\zeta_c\sim 4\times10^{15}$\,rad/s. 
Lower panel: Perturbative parameter $\beta_i$ 
for $\textrm{BP}$ and $\textrm{2D-P}$ (shown for all three faces of $\textrm{BP}$ crystal).}%
\label{diel}%
\end{figure}

The reversal in the rotation of the torque as a function of separation distance has never been reported in the literature to our knowledge. The sign-reversal in the torque reported here is above and beyond the well understood change in the sign arising from the periodic oscillatory dependence in the angle $\theta$\,\cite{Barash:1977ba,Munday:2005tbp,*Munday:20085errata,Somers:2017rcf,Somers:2017prl,Parsegian:1971anr,Lu:2016ml}.
We show that both black phosphorus ($\textrm{BP}$) and its two-dimensional (2D) monolayer 
phosphorene ($\textrm{2D-P}$) are suitable materials that permit $\beta_i(\omega_c)=0$, characterized 
by crossings in the plots of their in-planar components of the dielectric functions with 
respect to frequency, as shown in Fig.~\ref{diel}. We verify and report our confirmation 
of the separation distance dependent sign-reversal of the torque experienced in slabs 
comprising of $\textrm{BP}$ and $\textrm{2D-P}$. 
The critical distance $a_c$ is approximately $40$\,nm for $\textrm{BP}$ and $\textrm{2D-P}$,  
but with a little bit of material engineering it should be possible to construct materials
that suit a specific need.
In Fig.~\ref{tor-main} we showcase the result for $\textrm{BP}$, where 
two slabs
try to align their principal axes of polarizabilities in one direction below $a_c$, while above $a_c$ they try to align their principal axes perpendicular relative to their previous minimum energy orientation. This sign-reversal behavior is absent in the corresponding nonretarded torque.

The major stumbling block in the calculation of the torque
is that the electromagnetic modes in the presence of a biaxially polarizable
material do not separate into transverse electric (TE) and traverse magnetic (TM) modes. In our method, we circumvent this hindrance by choosing the system to be uniaxial in the absence of perturbation, $\beta_i=0$ for all $\omega$. 
For the setup shown in the inset of Fig.~\ref{tor-main}, we consider that the dielectric functions $\bm\varepsilon_i(\omega)$ of two nonmagnetic biaxially anisotropic slabs of thickness $d_i$ are diagonal in the basis of 
their principal axes ($\hat{\bf u}_i$, $\hat{\bf v}_i$, $\hat{\bf n}$),
\begin{equation}
{\bm\varepsilon}_i(\omega) \!=\!
\varepsilon^{u}_i(\omega) \,\hat{\bf u}_i \hat{\bf u}_i
+ \varepsilon^{v}_i(\omega) \,\hat{\bf v}_i \hat{\bf v}_i
+ \varepsilon^{n}_i(\omega) \,\hat{\bf n} \hat{\bf n},
\label{aniso-diel}%
\end{equation} 
where we chose one of the principal axis of each material to align along $\hat{\bf n}$, which is normal to the slabs as shown in Fig.~\ref{tor-main}.  
We earlier developed a perturbation method in~\cite{gear1,*gear2,*gear3} to study lateral Casimir 
forces arising due the asymmetry in the geometry of the system. Here, we extend our methods 
to incorporate asymmetry in the polarization. 
We decompose $\bm\varepsilon_i(\omega)$  of the biaxial materials as
\begin{equation}
{\bm\varepsilon}_i(\omega) = {\overline{\bm\varepsilon}}_i(\omega) 
+ \Delta{\bm\varepsilon}_i(\omega),
\label{ep-dec}
\end{equation}
where ${\overline {\bm\varepsilon}}_i(\omega)$ represents the uniaxial 
background 
\begin{equation}
{\overline {\bm\varepsilon}}_i(\omega)=\varepsilon^{\scriptscriptstyle \perp}_i(\omega)
\,(\hat{\bf u}_i \hat{\bf u}_i +\hat{\bf v}_i \hat{\bf v}_i)+\varepsilon^n_i(\omega) \,\hat{\bf n} \hat{\bf n}
\label{iso-uni}
\end{equation}
for which a closed form solution for the Green dyadic can be obtained. We define $\varepsilon^{\scriptscriptstyle \perp}_i(\omega)\!=\!(\varepsilon^{u}_i (\omega)+\varepsilon^{v}_i(\omega))/2 $ so that the degree of anisotropy that characterizes the biaxial nature of the material is then
completely captured inside
\begin{equation}
\Delta{\bm\varepsilon}_i(\omega) = 
\left(\frac{\varepsilon^{u}_i (\omega)-\varepsilon^{v}_i(\omega)}{2}\right) 
\,(\hat{\bf u}_i \hat{\bf u}_i -\hat{\bf v}_i \hat{\bf v}_i).
\end{equation}
This particular choice renders $\text{Tr} \, \Delta{\bm\varepsilon}_i(\omega) = 0$. The perturbative parameter in Eq.~(\ref{beta}) can now be written as $\beta_i(\omega)=\Delta\varepsilon_i(\omega)/\varepsilon^{\scriptscriptstyle \perp}_i(\omega)$, which implies ${\bm\varepsilon}_i(\omega)\! =\! {\overline{\bm\varepsilon}}_i(\omega) (1+\beta_i(\omega))$.
It is easy to find materials with small $\beta_i$ for all $\omega$. For example, for the plane of $\textrm{2D-P}$ and the corresponding face of the BP crystal the magnitude of the static values of the perturbative parameter, $\beta_i(0)$, are approximately $0.06$ and $0.09$, respectively (see Fig.~\ref{diel}).

We evaluate the contribution to the interaction energy from the terms that are the first-order in the perturbative parameter $\beta_i$, separately, to be zero. Thus, the leading-order contribution to the interaction energy at finite temperature $T$ is from the second-order term containing $\beta_1\beta_2$
\begin{equation}
{\cal E}^{(2)}(a,\theta)\!
=\! -\frac{k_BT  \cos 2\theta}{4 \pi }\!
\sum_{m=0}^{\infty}{'}\! \beta_{1}\! \beta_{2}\! \!\int_0^\infty\!\!\!\!\! k\, \mathrm{d}k\,
\textrm{Tr}(\widetilde{\bm R}_1\!\widetilde{\bm R}_2) e^{-2\kappa a},
\label{E2-def}
\end{equation}
where $k_B$ is the Boltzmann constant, $\theta$ is the angle between the in-planar principal axes of the two materials shown in the co-ordinate system in Fig.~\ref{tor-main}, and $k$ is the wave
vector perpendicular to the $\hat{\bf n}$ direction. The prime on the summation denotes that the zero frequency mode is taken with the half weight. We define $\kappa = \sqrt{k^2 +\zeta_m^2/c^2}$, where $\zeta_m$ is the imaginary Matsubara frequency defined in the caption of Fig.~\ref{diel}. 
The expression for the ``reduced reflection'' coefficient $\widetilde{\bm R}_i$ is given in the supplemental material~\cite{SM}. We highlight that an exact expression for interaction energy between two biaxially polarizable materials, that includes retardation, remains an open problem--Barash's result is for uniaxial materials suitably rotated to calculate the free energy. Thus, our approximate expression for the interaction energy in Eq.~(\ref{E2-def}) in the perturbative parameter $\beta_1\beta_2$, is a significant progress for the analysis of the interaction between biaxial materials. The details of our perturbation theory, which can be extended to higher orders and generalizes our earlier work~\cite{gear1} to include the total interaction energy will be presented elsewhere.


The leading order contribution to the torque per unit area on the dielectric slab is given by
\begin{equation}
{\mathcal T}^{(2)}(a,\theta) =- \frac{\partial}{\partial\theta} {\cal E}^{(2)}(a,\theta),
\label{tor-def}
\end{equation}
which replaces $\cos2\theta$ by $2\sin2\theta$ in Eq.~(\ref{E2-def}). In the nonretarded limit, when we take $d_i\to\infty$ and set $\varepsilon^v=\varepsilon^n$, we reproduce Barash's uniaxial result in the corresponding weak limit. The $2\theta$ 
dependence is a signature of bi-directional nature of fluctuation dependent 
polarizabilities. The dependence of torque on $a$ is of the form $1/a^2$ times a function of $d_i/a$, which is usually monotonic. In this article, we construct configurations of anisotropic materials that not only break away from the monotonous dependence on $a$ but also change sign by carefully selecting configurations such that $\beta_i(\zeta_c)=0$ for at least one frequency. (Note that $\beta_i$ will approach $0$ at high frequencies.)

For two identical materials, the torque displays a monotonic behavior as a function of $a$ because $\beta_1\beta_2=\beta_1^2\ge 0$ for all frequencies. However, for nonidentical materials, if $\beta=0$ at a characteristic frequency $\zeta_c$ for one of the interacting materials (or $\beta_i$ is zero for both the materials but at significantly different characteristic frequencies), then the Matsubara frequency modes above and below the characteristic frequency  will give contributions with opposite signs to the torque. The cancellation between the positive and negative contributions to the torque summed over all Matsubara frequencies in Eq.~(\ref{E2-def}) will decide the overall sign of the torque at a fixed separation distance. The contributions from the higher Matsubara frequencies dominate at short separation distances while lower frequencies are more important at larger separation distances. These two competing effects create a scenario where the torque between two materials 
can reverse its rotational sense as a function of the separation distance. The sign-reversal of the torque is 
a generic behavior for any set of materials with the aforementioned material properties and is independent of the relative orientation $\theta$. 

It is also relevant to note that the next-to-the-leading-order term ${\mathcal T}^{(4)}$ in the expression for the torque will contain $(\beta_1\beta_2)^2$ and $e^{-4\kappa a}$. Thus, this term is not only suppressed by the small magnitude of $\beta_i$ but also subdued exponentially compared to the second-order term ${\mathcal T}^{(2)}$. Thus, the inclusion of this term may affect the magnitude of the torque slightly but cannot affect the sign-reversal behavior of the torque.

\begin{figure}[t]
\includegraphics[width=8.5cm]{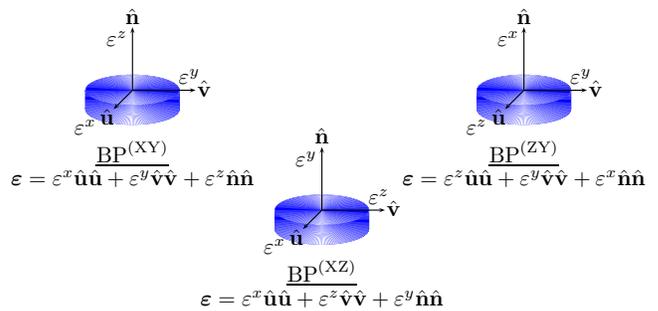}
\caption{Schematics for choices of interacting media for $\textrm{BP}$. We denote $\textrm{BP}^\textrm{(XY)}$ for the configuration when X-Y face of the $\textrm{BP}$ aligns perpendicular to $\hat{\bf n}$. Similarly, $\textrm{BP}^\textrm{(XZ)}$ and $\textrm{BP}^\textrm{(ZY)}$ describe the configurations when X-Z and Y-Z faces of $\textrm{BP}$ are set perpendicular to the $\hat{\bf n}$, respectively.}
\label{bp-face-def}%
\end{figure}

To illustrate the above mentioned change in the direction of the torque, we use $\textrm{BP}$ and $\textrm{2D-P}$, which are biaxially polarizable due to their puckered non-planar honeycomb structures\,\cite{Liu:2014hhm}. The 
optical properties of $\textrm{BP}$ and $\textrm{2D-P}$ are computed using the Vienna Ab-initio Simulation Package (VASP). The optB88-vdW functional\,\cite{klimes:2010dft,*klimes:2011dft} is used for structural relaxation while the revised Heyd-Scuseria-Ernzerhof (HSE) screened functional\,\cite{Krukao:2006shf} is used for the dielectric function calculations. The computed band gap 
energies of $\textrm{BP}$ and $\textrm{2D-P}$ are 0.38 and 1.52 eV, respectively, which are consistent with the previously reported results\,\cite{Malyi:2017mps,*Wang:2015mbp}. (See supplemental material\,\cite{SM} for details.)
Figure~\ref{diel} 
displays the dielectric tensor components $\varepsilon^{x}$, $\varepsilon^{y}$, and $\varepsilon^{z}$, along the principal axes ($\hat{\bf x}_i$, $\hat{\bf y}_i$, $\hat{\bf z}$) of $\textrm{BP}$ and $\textrm{2D-P}$ crystals as a function of $\zeta_m$. The components $\varepsilon^{x}$ and $\varepsilon^{y}$ of the dielectric function of $\textrm{BP}$ and monolayer $\textrm{2D-P}$ cross approximately at $4\times 10^{15}$\,rad/s. 
The BP crystal has three faces--each with a different degree of anisotropy. In the Casimir-Lifshitz setup, shown in the inset of Fig.~\ref{tor-main}, we have the choice to align different faces 
perpendicular to $\hat{\bf n}$, as delineated in Fig.~\ref{bp-face-def}. The perturbative parameters corresponding to the three orientations of the $\textrm{BP}$ crystal, and $\textrm{2D-P}$, are presented in the lower panel of Fig.~\ref{diel}. 
Note that $\beta^\textrm{(XZ)}$ and $\beta^\textrm{(ZY)}$ for $\textrm{BP}$ are never zero, which equips us with a suitable set of dielectric function for one of the interacting materials to test our theoretical predictions.

In the supplemental material~\cite{SM}, we show the comparison of our perturbative result applied to the interaction between two uniaxial materials, whose optical axes are along $\hat{\bf u}_i$, with the exact theory\,\cite{Barash:1977ba}. We used a simulated dielectric function for the $\textrm{BP}$ generated by setting the components $\varepsilon^{y}=\varepsilon^{z}$. Our leading order perturbative results match within $5.5\%$ of the exact theory for the separation distances $1$-$100$\,nm. 
Thus, emboldened by the remarkable performance of the leading-order result for $\textrm{BP}$, we now proceed to test our main prediction of the sign-reversal of the torque. 
\begin{figure}[t]
\includegraphics[width=8.5cm]{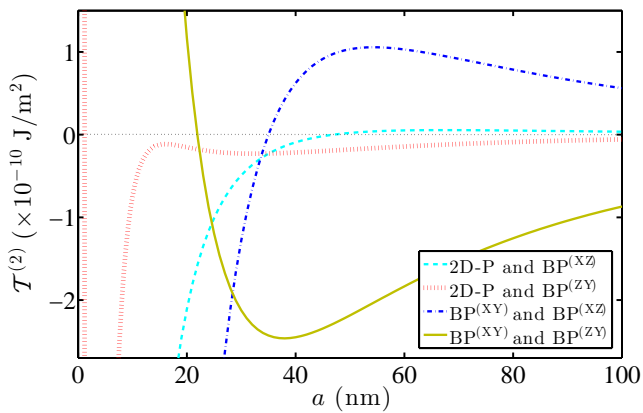}
\caption{Torque per unit area as a function of the separation distance $a$ between different combinations of $\textrm{BP}$ and $\textrm{2D-P}$ for $\theta=\pi/4$. The torque is not only nonmonotonic but also changes its sense of rotation as a function of $a$.}
\label{torque}%
\end{figure}
We evaluate the torque for the four different combinations of $\textrm{2D-P}$ and $\textrm{BP}^\textrm{(XY)}$ interacting with $\textrm{BP}^\textrm{(XZ)}$ and $\textrm{BP}^\textrm{(ZY)}$, which have a possibility of showing sign-reversal according to Eqs.~(\ref{E2-def}) and (\ref{tor-def}). Our results for the leading-order torque as a function of the separation distance $a$ are presented in Fig.~\ref{torque}. 
All the four cases show reversal in the direction of the torque at short 
separation distances ranging from $20$-$50$\,nm with the exception of interaction between $\textrm{2D-P}$ and  $\textrm{BP}^\textrm{(ZY)}$ in which the torque  
changes sign at a very short distance but then shows multiple extrema as a function of $a$. This suggests that by manipulating the dielectric properties one could find 
multiple separation distances where the torque acting between two anisotropic materials may change its rotational sense for any arbitrary orientation between their in-planar principal axes. The results presented in the figure are for $\theta=\pi/4$. A different value of the relative orientation $\theta$ will change the magnitude and sign of the torque with a periodicity of $\sin2\theta$.

\begin{figure}[t]
\includegraphics[width=8.5cm]{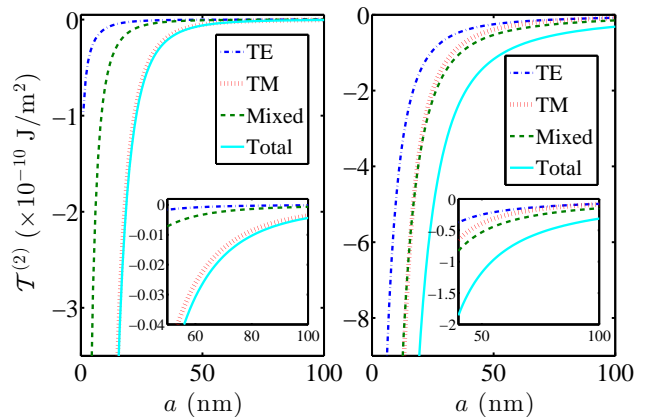}
\caption{Torque per unit area as a function of the separation 
distance $a$ between two identical $\textrm{2D-P}$ layers (left panel) and two identical 
semi-infinite $\textrm{BP}$ slabs (right panel). We show the TE, TM and mixed mode 
contributions, which have different hierarchical order in the two cases. }%
\label{modes}
\end{figure}

Next, we investigate the scenario when the two interacting media are identical. 
As mentioned earlier, the torque is monotonic in this situation. Figure~\ref{modes} shows the leading-order torque as a function of $a$ for the interaction between two identical $\textrm{2D-P}$ in the left panel and between two semi-infinite slabs of $\textrm{BP}^\textrm{(XY)}$ in the right panel. In contrast to that of 2D materials, the magnitude of the torque between thick media is bigger by one to two orders of magnitudes. A similar monotonic behavior appears if one of the materials is $\textrm{2D-P}$ and the other material is the $\textrm{BP}^\textrm{(XY)}$, as $\beta_i=0$ for both the materials at very close characteristic frequencies. 
Further, although the electromagnetic modes do not separate in biaxial systems, in our perturbation theory we can identify contributions to the torque from TE, TM, and a mixed mode, that depends on both TE and TM reflection coefficients defined for the uniaxial background of Eq.~(\ref{iso-uni}). 
The contributions from the TM mode to the torque dominates for small $a$ of about 10\,nm, followed by the mixed mode, with TE mode being negligible. All the modes begin to contribute comparably at large $a$ of about 100\,nm, but keep their hierarchical order for the case of interaction between two $\textrm{2D-P}$ layers. 
The interaction between two $\textrm{BP}$, on the other hand, presents a curious feature where the mixed mode over takes the TM mode at about $25$\,nm with TE mode also crossing over near $100$\,nm. From a fundamental point of view, this difference indicates a non-additive nature of the interlayer interaction in BP--also mentioned in Ref.\,\cite{Shulenburger:2015bp} in the context of interlayer binding energy.

Before closing our discussion, we qualitatively comment on the feasibility of measuring the effects discussed here. The experimental verification of the Casimir-Lifshitz torque has remained elusive 
due to the smallness of the magnitude of the torque~\cite{Capasso2011}. For a material slab with cross-sectional 
area of $100\,\mu$m$^2$, the torque is of the order of $10^{-20 }$\,Nm for $\textrm{BP}$ and  $\textrm{2D-P}$, which is in accordance with the other calculations reported in the literature~\cite{Munday:2005tbp,*Munday:20085errata,*Romanowsky:2008hac,*Somers:2017rcf,Somers:2017prl,Hu:2014book}. However, the retardation effect and using an intervening liquid medium have shown an appreciable change in the magnitude of the torque\,\cite{Somers:2017prl,Morgado:2013nws}. Other methods, like carrier injection to manipulate the dielectric functions, could provide enhancement in the torque. Newer experimental techniques as suggested in\,\cite{Guerout:2015tnp,*Xu:2017onr}  could be explored. Our primary motivation, here, is to highlight the distance dependent sign-reversal in the Casimir-Lifshitz torque that should be kept in mind in the quest for an experimental verification. One may verify the existence of the torque, at least in principle, utilizing the sign-reversal of the torque. Assume, for instance,  that the two planar materials at their initial positions  are  illuminated by 
a laser beam from above, and the scattering  pattern is observed. With the change of the separation distance between the slabs the torque will change its sense of rotation leading to a reorientation of the principal axes. Thus, the scattered radiation will have changed with the change of the separation distance. At least a qualitative change of the radiation pattern would be sufficient to verify the existence of the torque. The effect is analogus to Mie scattering from a dielectric sphere if  subjected to slight surface deformations.

In conclusion, non-separability of TE and TM modes continues to be a hindrance in finding an exact solution for the torque between two biaxially anisotropic materials--Barash's solution in Ref.\,\cite{Barash:1977ba} was for a uniaxial material. However, we have developed a perturbative method that overcomes this shortcoming to an excellent accuracy. The change in the reversal of the torque highlighted here is expected to be prevalent in materials, and is an open door for device engineering.

{\it Acknowledgements}\,-
We acknowledge support from the Research Council of Norway (Project No. 250346) and access to high-performance computing resources via SNIC and NOTUR. The work of KAM is supported in part by the US National Science Foundation (Grant No. 1707511).

\bibliography{biblio/b20141025-biaxial-anisotropy}

\end{document}